# Effect of Quantum correlation on Perfect K pair Quantum Network Coding

M. Mahdian[1] , R. Bayramzadeh[1]

[1]Faculty of Physics, Theoretical and Astrophysics Department ,University of Tabriz, 51665-163, Tabriz, Iran.

We find a protocol transmitting K quantum states crossly in the quantum networks with K pair sender-receiver and one bottleneck channels only with sharing quantum correlations between senders. In these networks senders want to multicast perfect quantum states to receivers. Perfect send on the quantum networks is impossible just using some additional resources like quantum correlation. This protocol has been used maximum entangled states and Quantum Discord which is best indicator of the quantum nature of the correlation to sharing between senders. At first, to send perfect K states in quantum network applied some maximum entanglement states to teleport quantum states, Enistein – Podolsky – Rosen (EPR) states as simple and two qubit states and Grrenberger – Horne – Zeilinger (GHZ) states and W states as kinds of three qubit states. Then reached that there is no difference between sharing these maximum entanglement states and must send same amount of classical bits in bottleneck channel. In the second part of this paper of sharing Quantum Correlation (quantum discord) perfect send of two quantum states is possible in the Butterfly network with fidelity equal one.

## I. Introduction

The excellent idea of network coding proposed by Ahlswede, et al [1] in 2000, opened up a new communication method of transmition information through networks. The network itself is given as a weighted, directed acyclic graph with the weights denoting the capacities of the edges with noiseless channels.Maximizing information exchange over classical communication networks has been a bigger subject among both the information theory and the networking societies. Especially, a large scale network often has a bottleneck point that causes a transmission rate relatively small for the size of its communication resource, Bottleneck restricts the total performance of communication. The important idea is to allow coding and replication of information locally at any intermediate node of the network for instance, this allows to send two bits simultaneously between two source-target pairs over several networks for which the same task cannot be solved by routing in the Butterfly network depicted in Fig. (1),then must used swap operation where a coding scheme is a choice of operations for internal nodes in graph.

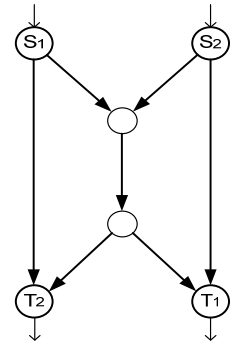

*Figure 1 :The Classical 2 pair Butterfly network that total flow going through each edge in the network does not exceed its capacity.*

Network communication was traditionally done by routing also known as the store and forward method in which received data is simply copied and forwarded without data processing, network coding in general offers throughput benefits as compared to the routing.Also network coding makes the upper bound on the achievable rate given by the min-cut max-flow condition is in fact always achievable.In other words, network coding allows one to send ***m***massages if and only if the value of any cut between the source node and each target node is at

E-mail: Mahdian@tabrizu.ac.ir

least *m* then network coding are good for high communication rate transmission. We refer to Refs [2-5] for extensive treatment of the classical network coding.The multicast problem is a task that can be elegantly solved by network coding; in this K pair problem all messages at one source node must be sent to each of several target nodes;For each $i \epsilon \{1…, K\}$ a massage $x_i$ is given at the source $S_i$, and has to be sent to the target $t_i$ through the network under the condition that each edge has unit capacity, The Butterfly network is an instance of the two pair problem. Note that the classical K pair problem, that the channels and messages are classical, is one of the most important network coding problems [2].

Since the faults of classical mechanics its must replaced with general quantum mechanics but making the copy of digital information is convenient but it is not possible exactly for quantum information. The important question is whether or not quantum network coding is possible, we consider a problem of quantum communication between parties that are connected through a network of quantum channels. In spite of no-cloning theorem and has not quantum copy operation, we can send quantum information in quantum networks that have quantum channels and quantum state transmit in long distance using network coding to reduce the amount of quantum communication. At 2007 Hayashi , et al [11] performed network coding discuss to quantum systems and their resulted are that perfect quantum network coding is impossible and we can't send quantum states in quantum networks with fidelities equal one. They concluded that we can achieve fidelities greater than 1/2 and lower that 1. For more studying on quantum network coding [6, 11-13] discusses about quantum transmission on quantum networks with quantum channels. In the quantum case, the methods used in the classical case cannot be applied directly and must used additional resources for example classical communications and entanglement [6,13]. The K pair problem is one of the most important network coding problem, and in quantum discussion quantum K pair problem is not perfectly solvable, we motivated with this papers to investigate K pair quantum network coding. Perfect network coding is impossible for all quantum protocols, the fidelities at nodes $t_1$ and $t_2$ are <1; without additional resources whereas imperfect network coding is possible: there exists a quantum protocol whose fidelities at targets are >1/2. To reduce the amount of classical communication as much as possible, the work of Leung, et al [6], they investigated various settings of quantum network coding assisted with supplied resources such as free classical communication or entanglement. They considered the case where classical communication can be sent only between each pair of nodes connected by a quantum channel and only in the direction of that quantum channel. They found that quantum two pair problem on the butterfly network cannot be solved even under this model.In 2009, under the assumption that there was no prior entanglement shared among any of the parties, but that classical communication was free, Kobayashi et al, gave a perfect quantum network communication protocol based on classical network coding and because of classical communication does not increase the amount of entanglement of the system [7-9].This protocol requires only one qubit transmission or two bits transmission in each channel in the network.The single-qubit quantum and 2 bit classical channels are mutually in equivalent resources. Note that super dense coding implies that a single qubit quantum channel and shared 1 ebit entanglement together have the capacity of 2 bit classical channels, and teleportation shows 2 bit classical channel and shared 1 ebit entanglement together have the capacity of a single qubit quantum channel. Other additional resources are entanglement and Prior entanglement provides some miracle performances in quantum information. Moreover, prior entanglement enables the transmission of quantum state only by sending classical information. Hayashi finded a protocol to sending perfect states in the butterfly network with prior entanglement among senders [13].

In real situations, however, it is most of the time not possible to have a maximum entangled state at one's disposal. Because of the interaction with the environment, the quantum state of any system will became the mixed state after a certain period then always maximum entanglement states is not accessible, entanglement is susceptible to interaction with the environment which results in disentanglement. So our group wants to study quantum network coding with quantum correlation that are more interesting and powerful than entanglement [18]. Entanglement may disappear completely after a finite time, an effect known as entanglement sudden death. There exist, however, quantum correlations more general and more fundamental than entanglement.Resistance of quantum correlation against decoherency is more

E-mail: Mahdian@tabrizu.ac.ir

than entangled states and we have high speed transition with quantum correlation. There are many application of quantum correlation in quantum information theory,one of the usage is quantum teleportation [14, 15] such also used in this paper. One of the other quantum correlation types which is more general than entanglement, is quantum discord. So quantum discord , a more robust and easy to gain access to phenomenon than entanglement, can also deliver a quantum advantage introduced by Olivier and Zurek [16] for a rather limited set of bipartite systems that has nonzero value for some mixed separable state and as an informational-theoretical measure of quantumness of correlations. Quantum discord is defined as the difference between general mutual information and classical information [16-18].

This paper investigates the K pair quantum network coding and sending perfect quantum states on the networks. This paper used *K* maximum entanglement states to sharing between senders as additional resources to remove limitation on quantum information processing. Using results of teleportation *K* quantum states and amount of classical communication, we concluded for quantum network coding must send same amount of classical bits. As the results, find that same amount of classical communication needed for these maximum entanglement states; EPR states, GHZ states and W states; that calculated 2(K-1) bits on bottleneck channel for crossly sending of course this protocol alloweither qubits or bits classical communication in networks. Main discussion of the paper is thatin the quantum network if senders want to send some quantum states with special angles in the Bloch sphere to receivers, fidelity one can approached with quantum discord.

This paper has organized as follows: In Sec.2 we described K pair perfect quantum network coding with sharing maximum entanglement EPR states, In Sec.3 we review K pair Quantum network coding with sharing GHZ and W states, in Sec. 4 we introduces the Quantum correlation and described how Quantum teleportation and Quantum network coding can be perfect with using quantum discord, Conclusions are then presented in Sec.5.

## II. Kpair Quantum network coding with sharing Maximum entanglement EPR states

Entanglement is considered to be a fundamental resources of quantum information processing such as

E-mail: Mahdian@tabrizu.ac.ir

quantum cryptography, quantum teleportation, and quantum computation and so on. Entangled states, which are called quantum channels in teleportation processing, make it possible to send an unknown quantum state in long distance that maximum entangled states, i.e. Bell states, are used as quantum channels in perfect quantum teleportation. As additional resources to the quantum network coding we allow sharing maximum entanglement states between senders to send perfect quantum states.If two bit transmission is allowed instead of one qubit transmission, only using quantum teleportation, perfect transmission of quantum state is available. Therefore perfect state transfer is possible, we also want to achieve it with as few uses of the network as possible then must used Network coding.We have generalized quantum 2 pair perfect quantum network coding with sharing two maximum entanglement EPRstates [13, 15] in Eq. (1) to *K*pair quantum network coding and calculated the capacity of bottleneck channel to send classical communications.

$$\left|\phi^{EPR}\right\rangle = \frac{1}{\sqrt{2}}\left(\left|00\right\rangle + \left|11\right\rangle\right) \quad (1)$$

For teleportation two quantum states [14] in the quantum butterfly network for sending perfect states showed in Fig. (1), we need two classical bits and must send two bits in bottleneck channel to decoding arbitrary states in targets according to [13,15].

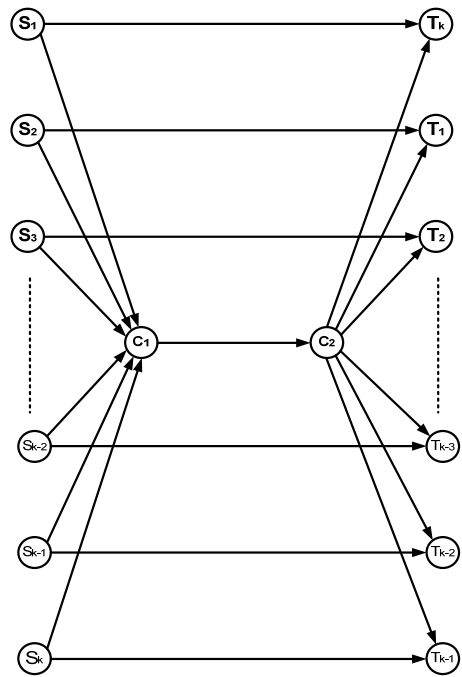

*Figure 2: K pair quantum network that it's possible perfect send with sharing K maximum entangled states.*

K pair problem which have *K* sources and *K* targets in Fig. (2) Must send *K* quantum states of specific sources to their targets. For simplicity, first we treat 3 pair quantum network and presented its quantum circuit in Fig. (3) To sending perfect three states with sharing three *EPR* entangled states.

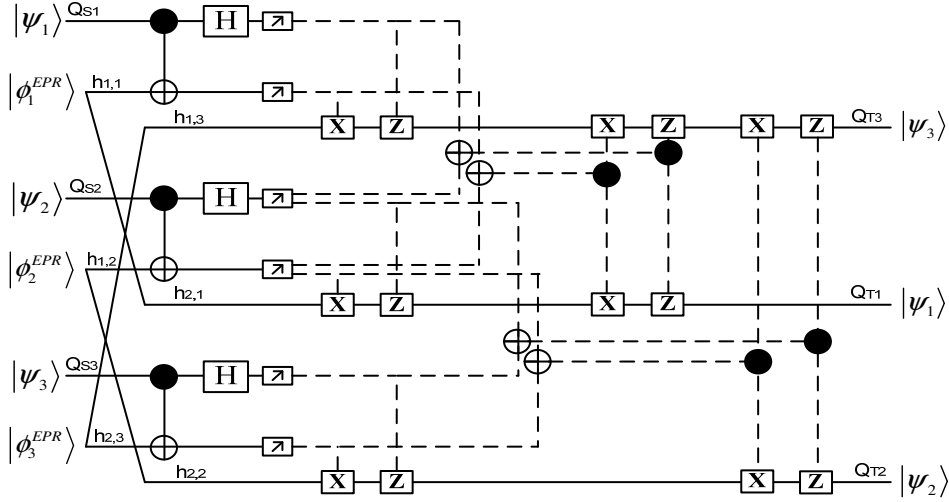

*Figure 3: The Quantum circuit for perfect 3 pair quantum network with sharing three maximum entangled EPR states.*

Now, we have initial state according to Eq. (2) which every $|\psi_i\rangle$ represent the information of $i^{th}$ sender, and the goal is send this state perfectly.

*initial state* :

$$|\psi_1\rangle \otimes |\psi_2\rangle \otimes |\psi_3\rangle \quad (2)$$

Then shared three maximum entangled states between senders and send last qubit of entanglement states or state of Bob in Teleportation processing with sender which we want to send crossly arbitrary quantum states to proper receivers, from this point Eq. (3) duplicated the quantum state.

$$\begin{aligned}
&|\psi_1\rangle \otimes |\psi_2\rangle \otimes |\psi_3\rangle \\
&\otimes |\phi_1^{EPR}\rangle \otimes |\phi_2^{EPR}\rangle \otimes |\phi_3^{EPR}\rangle \\
&\equiv |Q_{S_1}\rangle \otimes |Q_{S_2}\rangle \otimes |Q_{S_3}\rangle \\
&\otimes (|h_{1,1}, h_{2,1}\rangle + |h_{1,1}, h_{2,1}\rangle) \\
&\otimes (|h_{1,2}, h_{2,2}\rangle + |h_{1,2}, h_{2,2}\rangle) \\
&\otimes (|h_{1,3}, h_{2,3}\rangle + |h_{1,3}, h_{2,3}\rangle) \\
&= (\alpha_1|0\rangle + \beta_1|1\rangle) \\
&\otimes (\alpha_2|0\rangle + \beta_2|1\rangle) \\
&\otimes (\alpha_3|0\rangle + \beta_3|1\rangle) \\
&\otimes (|00\rangle + |11\rangle) \\
&\otimes (|00\rangle + |11\rangle) \\
&\otimes (|00\rangle + |11\rangle)
\end{aligned} \quad (3)$$

The replaced form of quantum states multiply in Eq. (4) is the same with Eq. (3) for example $|000\rangle_{S_1 S_2 S_3}$ described three multiple states belong to sources that all of them were zero.

$$\begin{aligned}
&\big(\alpha_1\alpha_2\alpha_3|000\rangle_{(S_1,S_2,S_3)} + \alpha_1\alpha_2\beta_3|001\rangle_{(S_1,S_2,S_3)} \\
&+ \alpha_1\beta_2\alpha_3|010\rangle_{(S_1,S_2,S_3)} + \alpha_1\beta_2\beta_3|011\rangle_{(S_1,S_2,S_3)} \\
&+ \beta_1\alpha_2\alpha_3|100\rangle_{(S_1,S_2,S_3)} + \beta_1\beta_2\alpha_3|110\rangle_{(S_1,S_2,S_3)} \\
&+ \beta_1\alpha_2\beta_3|101\rangle_{(S_1,S_2,S_3)} + \beta_1\beta_2\beta_3|111\rangle_{(S_1,S_2,S_3)}\big) \\
&\otimes \big(|00\rangle_{h_{1,1},h_{2,1}} |00\rangle_{h_{1,2},h_{2,2}} |00\rangle_{h_{1,3},h_{2,3}} \\
&+ |00\rangle_{h_{1,1},h_{2,1}} |00\rangle_{h_{1,2},h_{2,2}} |11\rangle_{h_{1,3},h_{2,3}} \\
&+ |00\rangle_{h_{1,1},h_{2,1}} |11\rangle_{h_{1,2},h_{2,2}} |00\rangle_{h_{1,3},h_{2,3}} \\
&+ |11\rangle_{h_{1,1},h_{2,1}} |00\rangle_{h_{1,2},h_{2,2}} |00\rangle_{h_{1,3},h_{2,3}} \\
&+ |00\rangle_{h_{1,1},h_{2,1}} |11\rangle_{h_{1,2},h_{2,2}} |11\rangle_{h_{1,3},h_{2,3}} \\
&+ |11\rangle_{h_{1,1},h_{2,1}} |11\rangle_{h_{1,2},h_{2,2}} |00\rangle_{h_{1,3},h_{2,3}} \\
&+ |11\rangle_{h_{1,1},h_{2,1}} |00\rangle_{h_{1,2},h_{2,2}} |11\rangle_{h_{1,3},h_{2,3}} \\
&+ |11\rangle_{h_{1,1},h_{2,1}} |11\rangle_{h_{1,2},h_{2,2}} |11\rangle_{h_{1,3},h_{2,3}}\big)
\end{aligned} \quad (4)$$

For perform protocol to perfect send in 3 pair quantum network, apply the *Hadamard* gate on quantum states of three sources states and *CNOT* gate to proper states Eq. (5) which these gates act according to Eq. (6),

$$H[Q_{S_1}, Q_{S_2}, Q_{S_3}]$$

$$CNOT\,[(Q_{S_1}, h_{1,1}), (Q_{S_2}, h_{1,2}), (Q_{S_3}, h_{2,3})] \quad (5)$$

$$H = \frac{1}{\sqrt{2}} \sum_{x,z \in F_2} (-1)^{xz} |z\rangle\langle x|$$

$$CNOT^{(1,2)} = \sum_{x,y \in F_2} |x\rangle_1 |x+y\rangle_2 \langle x|_1 \langle y|_2$$

$$\sigma_x = \sum_{x \in F_2} |x+1\rangle\langle x|$$

$$\sigma_z = \sum_{z \in F_2} (-1)^z |z\rangle\langle z| \quad (6)$$

For *EPR* states sharing on the teleportation processing there are four measurement basis that is same for network coding to access classical bits for sending to targets,

$$\text{basis of measurement}: |00\rangle, |01\rangle, |10\rangle, |11\rangle \quad (7)$$

According to Eq. (6) after earn two classical bits of each sources and applying *X* and *Z* Pauli transformation to remove additional phases on Bob qubit, those bits must send to internal nodes in the network and because of limitation on the capacity of bottleneck channel, encoding of classical bits is necessary.

After applying *CNOT* and *Hadamard* transformation the Alice states must be measured on suitable basis. The attained classical bits that are two quantity 0 and 1. For recovering perfect states on the last qubit of entanglement states must a proper *X* and *Z* gates applied. At the end of using classical coding the targets can access final state such as Eq. (8), that is equal to initial state which every $|\psi_i\rangle$ represent the information of $i^{th}$ targets.

$$\begin{aligned}\text{final state}: \\ |\psi_1\rangle \otimes |\psi_2\rangle \otimes |\psi_3\rangle \\ = |Q_{T_1}\rangle \otimes |Q_{T_2}\rangle \otimes |Q_{T_3}\rangle\end{aligned} \quad (8)$$

Interesting point is that for transverse information of the senders to the targets must Coding used to reduce classical mass of the communications. Assuming *(i, j),(k, l),(m, n)* are classical bits that obtained after

E-mail: Mahdian@tabrizu.ac.ir

measurement of the first, second and third quantum states. For example in Fig. (3), the state $h_{2,2}$ should be send second state to its receiver as same as teleportation must apply $X^l Z^k$ gates on this state to receive. But this state belongs to third of the Alice state and in this circuit $X^n Z^m$ transformation effected. Now, using of coding and summing received bits in module two on the internal nodes, perfect sending can be approached with sending two bits *n+l* and *m+k* on the bottleneck channel.

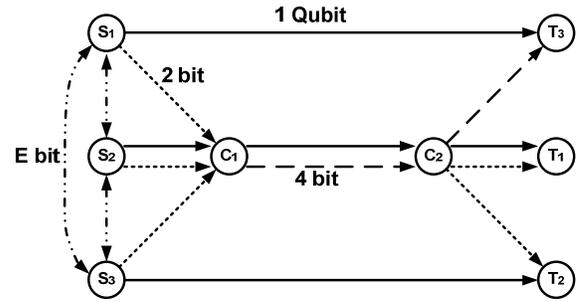

*Figure 4:* Capacity of channels in 3 pair quantum network for perfect sending three quantum states.

Therefore in 3 pair network to perfect send three quantum states to its receivers there should be sent 4 bits *(i+k, j+l, k+m, l+n)* on the bottleneck channel, this is represented in Fig. (4). It's possible to generalize 3 pair quantum network to K pair quantum network coding. This paper introduce a new generalization for *K* pair network that must send *2(K-1)* bits on bottleneck channel whereas K pair problem does not solved without network coding. If the ordering of receiver places be clockwise or non-clockwise and if there is a definite permutation in ordering of receiver places, for perfect transverse of K quantum states in quantum networks number of bits is needed are *2(K-1)*. We can define K as the place of K receivers which replaced in the network and any receivers do not be in front of its senders for perfect send then it needs *2(K-1)* bits on the bottleneck channel. If all receivers placed in front of its senders and one quantum channel supposed as pass channel for perfect send we don't need any classical bits whereas in these setting perfect quantum network needs K maximum entanglement *EPR* states.

## III. Quantum network coding with sharing GHZ states

Entanglement in three qubits is more complicated than that in two qubits. Recently, the entanglement of three qubits was classified by separable, bi-separable, *GHZ* and *W* states. The *GHZ* class state cannot be transformed to the W class by the local operation and classical communication. Although many suggestion has utilized the *GHZ* state in quantum teleportation and some proposals have suggested an implementation of the *W* state. The *GHZ* state and the *W* state can be used as the quantum channel for the perfect teleportation of an unknown qubitstate. In these type of teleportation and sending quantum states in quantum channels, Alice channel has three qubit channels but in *EPR* states sharing the Alice channel has two qubit channels. At the first look, decoding of any states in any receivers it's obvious that it need three bits indeed on the teleportation with sharing *GHZ* state that agreeswithEq. (9) Two bits are necessary for decoding and perfect sending and some otherqubits in the middleof the stateno longer is needed.

$$|\phi^{GHZ}\rangle = \frac{1}{\sqrt{2}}(|000\rangle + |111\rangle) \qquad (9)$$

For sending $K$ quantum states in the quantumnetworkwith sharing *GHZ* statespresented in Fig. (5). Therefore, there is no difference between *EPR* and *GHZ* states for sharing and in both of them it sends same amount of bits on bottleneck channels.If share $K$ states of EPR or *GHZ* states in $K$ pair network must send $2(K-1)$ bits to perfect send.

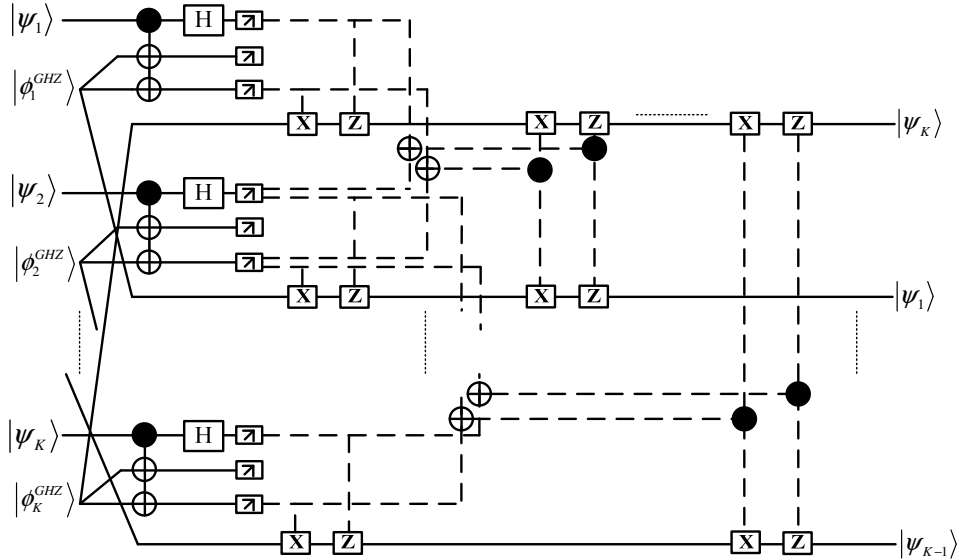

*Figure 5: The Quantum circuit for Perfect K pair quantum network sharing Ktimes of GHZ states.*

### A. Quantum network coding with sharing W states

The fact that *W* statein Eq. (10) having different type of *GHZ* states entanglement is an interesting topic for studying. The basis of measurement in teleportation with sharing *W* state are same as basis of *GHZ* state sharing then the classical bits needed for decoding and applying Pauli transformation for both of them are same.

$$|\phi^W\rangle = \frac{1}{\sqrt{3}}(|100\rangle + |010\rangle + |001\rangle) \qquad (10)$$

The method that how we share these states dependent to how we want to multicast quantum states to receivers.For instance, with supposition exist one quantum channel of one sender to another receiver directly with onequbit capacity quantum channel.If we want to send quantum states of one sender to its receiver thatexist in front of another sender, we share maximum entanglement states between these two senders. We attend to share three qubit states so every three qubit quantum states for teleportation have two parts that called the Alice and Bob channels.The two qubitof three qubit states belong to

E-mail: Mahdian@tabrizu.ac.ir

the Alice that it want to send her quantum state perfectly to Bob that have just one qubitchannel. To have perfect sending we must share Bob's qubit with definite sender. The quantum gates that needs for sending*K*quantum states with sharing *W* states as same as the quantum circuit presented in Fig. (5) Just these difference that must share *K*times of *W* states instead *GHZ* states and the bits needed in the bottleneck channel were equal.

## IV. Quantum Network Coding using Quantum Discord

### A. Introduction of Quantum Correlation

Entanglement measure of concurrence and quantum discord are two different types of the quantum correlations. Several measures of quantum correlation have been investigated in the literature and among them the quantum discord has recently received a great deal of attention. The quantum discord measures quantum correlations of a more general type than entanglement; such as there existsseparable mixed states having nonzero discord. Interestingly, proved both theoretically and experimentally that such states provide computational speedup compared to classical states in some quantum computation models Ref. [19]. For a given quantum state ρ of a composite system *AB* the total amount of correlations, is quantified by the quantum mutual information defined as Eq. (11).

$$I_{tot}(\rho) = S(\rho_A) + S(\rho_B) - S(\rho) \qquad (11)$$

Where $S(\rho)$ denotes the von Neumann entropyaccording to Eq. (12),

$$S(\rho) = -Tr(\rho \log \rho) \qquad (12)$$

And $\rho_{A(B)}$ and $\rho$ are the reduced density matrix of subsystem *A(B)* and the density matrix of the total system, respectively where $\rho_A = Tr_B(\rho)$, $\rho_B = Tr_A(\rho)$.

An alternative version of the mutual information can be defined as Eq. (13),

$$\tilde{J}_A(\rho) = S(\rho_B) - min_{\{E_k^A\}} \sum_k p_k S(\rho_{B|k}). \qquad (13)$$

Where the minimum is taken over all possible *POVM*s $\{E_k^A\}$ on subsystem A with $p_k = Tr(E_k^A \rho)$ and $\rho_{B|k} = Tr_A(E_k^A \rho)/p_k$. Since $\tilde{J}_A(\rho)$ quantifies the classical correlation, the difference in Eq. (14),

$$\widetilde{D}_A(\rho) = I(\rho) - \tilde{J}_A(\rho) \qquad (14)$$

Defines the quantum discord that quantifies the quantum correlation. Also the minimum in Eq. (14) can be taken over all von Neuman measurement and we denote the corresponding classical correlation as $\tilde{J}_A(\rho)$ and quantum discord as $D_A(\rho)$, respectively Ref (20).

### B. Application of Quantum Correlation on the Teleportation process

To perform teleportation procedure, we must have a quantum state to be shared between Alice and Bob as quantum channel. Correlations in this quantum state, has an important role in teleportation is that the process won't be done without these correlations. We consider single qubit state which want to teleport as

$$|\psi_{in}\rangle = \cos\frac{\theta}{2}|0\rangle + e^{i\phi}\sin\frac{\theta}{2}|1\rangle \qquad (15)$$

$|\psi_{in}\rangle$ in Eq. (15),

Where $\theta$ and $\phi$ are real number define a point on the unit three dimensional Bloch sphere, it provides auseful means of visualizing the state of single qubit.For simplicity, we take as initial states of the composite system of channels a class of states with

$$\rho_c = \frac{1}{4}\left(I \otimes I + \sum_{i=1}^{3} c_i \sigma_i \otimes \sigma_i\right) \qquad (16)$$

maximally mixed marginal, as $\rho_c$ from Eq. (16),

Where $c_i$ is a real number such that $0 \leq |c_i| \leq 1$ for every i and I the identity operator of the total systemand $\sigma_i$ are and Pauli operators, this class of states includes the Bell states if $|c_1| = |c_2| = |c_3| = 1$.Quantum discord computed for two-qubit*X* states Ref. [18] with this, we obtain quantum discord for density matrix that is defined in Eq. (16).Thenwe want to see the effect of quantum discord on quantum correlation, so investigate the teleportation with

E-mail: Mahdian@tabrizu.ac.ir

$$F = \frac{1}{2}e^{-i\phi}(e^{2i\phi}\cos^2\frac{\theta}{2}((2-t_1\cos\theta)\cos\phi$$
$$-i\sin\phi(1+c_3(-1+2\cos\theta)))$$
$$-\sin^2\frac{\theta}{2}((-2-t_1\cos\theta)\cos\phi$$
$$+i\sin\phi(-1+c_3(1+2\cos\theta)))) \quad (22)$$

channel that its entanglement is zero but discord is non-zero.

For this purpose, use the separable condition *(Positive - Partial - Transpose)*Ref. [21], of quantum channel and obtain the following equation Eq. (17)

$$\rho_c = \frac{1}{4}\begin{pmatrix} 1+c_3 & 0 & 0 & 1-c_3 \\ 0 & 1-c_3 & 1+c_3 & 0 \\ 0 & 1+c_3 & 1-c_3 & 0 \\ 1-c_3 & 0 & 0 & 1+c_3 \end{pmatrix}$$
(18)

on $c_i$;

By substituting Eq. (17) in Eq.(16), we get the channel density matrix equal to Eq. (18).

If use the Eq. (18) as channel, the entanglement is zero, Ofcourse; we suppose that Alice wants to teleport state expressed as Eq. (15) to Bob while the channel described as Eq. (16). According to Eq. (15) and Eq. (18), we have output according to Eq. (19),

$$t_1 = 1-c_3, t_2 = 1+c_3 \quad (20)$$

$$F = \langle \psi_{in} | \rho_{out} | \psi_{in} \rangle \quad (21)$$

$$c_1 - c_2 = 1 - c_3, \ c_1 + c_2 = 1 + c_3 \quad (17)$$

$$\rho_{out} = \frac{1}{2}\left(\left[t_1\sin^2\left(\frac{\theta}{2}\right)+t_2\cos^2\left(\frac{\theta}{2}\right)\right]|0\rangle\langle 0|\right.$$
$$+\left[t_1 e^{-i\phi}\cos\left(\frac{\theta}{2}\right)\sin\left(\frac{\theta}{2}\right)+t_2 e^{i\phi}\cos\left(\frac{\theta}{2}\right)\sin\left(\frac{\theta}{2}\right)\right]|0\rangle\langle 1|$$
$$+\left[t_2 e^{-i\phi}\cos\left(\frac{\theta}{2}\right)\sin\left(\frac{\theta}{2}\right)+t_1 e^{i\phi}\cos\left(\frac{\theta}{2}\right)\sin\left(\frac{\theta}{2}\right)\right]|1\rangle\langle 0|$$
$$+\left.\left[t_2\sin^2\left(\frac{\theta}{2}\right)+t_1\cos^2\left(\frac{\theta}{2}\right)\right]|1\rangle\langle 1|\right) \quad (19)$$

Which for simplicity have Eq. (20);

To measure the overlap between input and output states there are quantity that is called Fidelity and can be given by Eq. (21),

And then the fidelity calculated according to Eq. (22),

By considering $\phi = 0$, we plotted the fidelity of quantum teleportation versus $\Theta$ and $\phi$ in Fig. (6). Noting that in this figure it is assumed that $c_3 = 0$. Fig. (7) plotted the fidelity as function of $\Theta$ such that $c_3 = \phi = 0$. This figure shows that the fidelity only at point $\Theta = \pi/2$ is equal to one and elsewhere is always smaller than one. So we can transfer one quantum state perfectly under certain condition by using the quantum discord.

E-mail: Mahdian@tabrizu.ac.ir

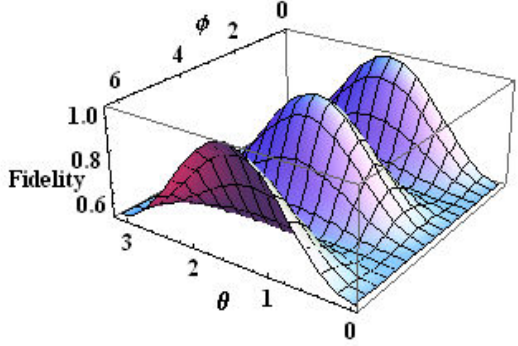

*Figure 6 : The Fidelity of quantum Teleportation as function of $\Theta$ and $\phi$ for $c_3=0$.*

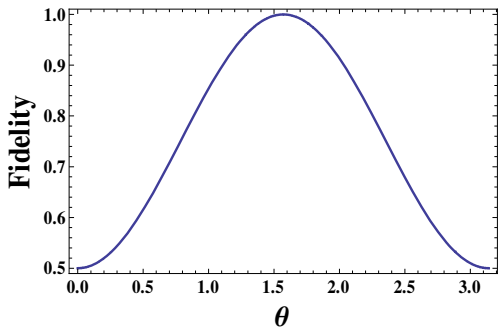

*Figure 7 : The Fidelity of quantum Teleportation as function of $\Theta$ for $\phi=0$ and $c_3=0$.*

At end we concluded that, both of quantum discord and entanglement played an important role in quantum teleportation and in absence of entanglement, discord causes non-zero fidelity.

### C. Application of Quantum Correlation in the Quant.um Network Coding

In the famous Butterfly network, we consider to sending two quantum states with sharing two quantum states that have not entanglement and using quantum discord as channels. In these setup, we have two quantum statessame as Eq. (23) that must send to receivers with different special angles,

$$|\psi_{1in}\rangle = \cos\frac{\theta_1}{2}|0\rangle + e^{i\phi_1}\sin\frac{\theta_1}{2}|1\rangle$$
$$|\psi_{2in}\rangle = \cos\frac{\theta_2}{2}|0\rangle + e^{i\phi_2}\sin\frac{\theta_2}{2}|1\rangle \quad (23)$$

And two quantum states that share between senders as quantum channels wrote in Eq. (24) which have quantum discord.

$$\rho_c = \frac{1}{4}\left(I \otimes I + \sum_{i=1}^{3} c_i \sigma_i \otimes \sigma_i\right)$$
$$\rho_{c'} = \frac{1}{4}\left(I \otimes I + \sum_{i=1}^{3} c'_i \sigma_i \otimes \sigma_i\right) \quad (24)$$

Using of results teleportation process with quantum discord, we can generalized to the butterfly network of two pair source- targets and send two quantum states perfectly of sources to targets. Then if quantum states belong to sources were special states with particular angles on Bloch sphere namely Eq. (25) perfect send accessible.

$$\begin{cases} \theta_1 = \theta_2 = \dfrac{\pi}{2}, \\ \phi_1 = \phi_2 = 0, \\ c_3 = c'_3 = 0. \end{cases} \quad (25)$$

This condition on transmission quantum states can applied to $K$ pair quantum networks that the goal is sending perfect $K$ quantum states of sources to receivers.

### V. Conclusion

It's possible to send $K$ quantum states of $K$ senders to $K$ receivers with fidelity equal one in quantum networks. In this paper have been used of quantum correlation to send perfect quantum states. Using of maximum entangled states; *EPR, GHZ* and *W* states we must used classical communication in bottleneck channels. In *K* pair networks must send *2(K-1)* times of bits on bottleneck channel for all three entangled states sharing if targets placed on the permutation places. Using of quantum correlation can send quantum states of senders to receivers that have not any prior entanglement.

E-mail: Mahdian@tabrizu.ac.ir

E-mail: Mahdian@tabrizu.ac.ir